\documentclass[twocolumn,showpacs,preprintnumbers,amsmath,amssymb]{revtex4}

\usepackage{graphicx}
\usepackage{dcolumn}
\usepackage{bm}
\usepackage[usenames]{color}

\begin{document}

\title{Quasi-Two-Dimensional Fermi Surfaces and Coherent Interlayer Transport in KFe$_2$As$_2$ }

\author{M. Kimata$^{1,6   \dagger}$, T. Terashima$^{1,6}$, N. Kurita$^{1,6}$, H. Satsukawa$^1$, A. Harada$^1$, K. Kodama$^{1,2}$, A. Sato$^{1,6}$, M. Imai$^{1,6}$, \\K. Kihou$^{3,6}$, C. H. Lee$^{3,6}$, H. Kito$^{3,6}$, H. Eisaki$^{3,6}$, A. Iyo$^{3,6}$, \\T. Saito$^{4}$, H. Fukazawa$^{4,6}$, Y. Kohori$^{4,6}$, H. Harima$^{5,6}$, and S. Uji$^{1,2,6}$
}
\affiliation{$^1$National Institute for Materials Science, Tsukuba, Ibaraki 305-0003, Japan}%
\affiliation{$^2$Graduate School of Pure and Applied Science, University of Tsukuba, Ibaraki 305-8577, Japan}
\affiliation{$^3$National Institute of Advanced Industrial Science and Technology (AIST), Tsukuba, Ibaraki 305-8568 Japan}
\affiliation{$^4$Department of Physics, Chiba University, Chiba 263-8522, Japan}
\affiliation{$^5$Department of Physics, Graduate School of Science, Kobe University, Kobe, Hyogo 657-8501 Japan}
\affiliation{$^6$JST, Transformative Research-Project on Iron Pnictides (TRIP), Chiyoda, Tokyo 102-0075, Japan}
\affiliation{$^{\dagger}$Present address: Institute for Solid State Physics, University of Tokyo, Kashiwa, Chiba 277-8581, Japan}


\date{\today}

\begin{abstract}
We report the results of the angular-dependent magnetoresistance oscillations (AMROs), which can determine the shape of bulk Fermi surfaces in quasi-two-dimensional (Q2D) systems, in a highly hole-doped Fe-based superconductor KFe$_2$As$_2$ with $T_c\approx$ 3.7 K. From the AMROs, we determined the two Q2D FSs with rounded-square cross sections, corresponding to 12$\%$ and 17$\%$ of the first Brillouin zone. The rounded-squared shape of the FS cross section is also confirmed by the analyses of the interlayer transport under in-plane fields. From the obtained FS shape, we infer the character of the 3d orbitals that contribute to the FSs. 
\end{abstract}

\maketitle

Superconducting mechanism of the newly discovered Fe-based superconductors (the maximum transition temperature $T_c$ of about 55 K) has attracted much attention in recent years\cite{kamihara,kito,ren}. The Fermi surface (FS) topology is an essential constraint on the possible mechanism, so that its determination has particular importance. Especially the bulk FS property is crucial because superconductivity is a bulk electronic state. Fe-based superconductors have two-dimensional (2D) square lattices of the Fe ions, and consequently, the quasi 2D electronic structures. According to the band calculations\cite{lebegue,shingh}, the conduction bands mainly consist of almost degenerate five $3d$ orbitals of the Fe ions: multiple FSs exist. Experimentally, considerable efforts to probe the FSs of the Fe-based superconductors have been made by various techniques, such as quantum oscillations (de Haas-van Alphen (dHvA) or Shubnikov-de Haas (SdH) effects) and angle-resolved photoemission spectroscopy (ARPES)\cite{coldea, lu, sato}. Quantum oscillation measurements enable us to probe the FSs accurately. However, the measurements cannot determine the shape of the FS cross section and the position in the momentum space. On the other hand, ARPES can determine the band structure and the FSs directly. However, ARPES is a surface probe, so that the results do not necessarily reflect the bulk properties of the material\cite{lu}. For low-dimensional systems, the angular-dependent magnetoresistance oscillation (AMRO) can also be a powerful tool to investigate the FSs\cite{kar,kajita,yamaji}. This method is a bulk probe, as well as the quantum oscillations, and can clearly determine the FS cross sectional shape of the Q2D FSs. 

\begin{figure}
\includegraphics[width=7cm]{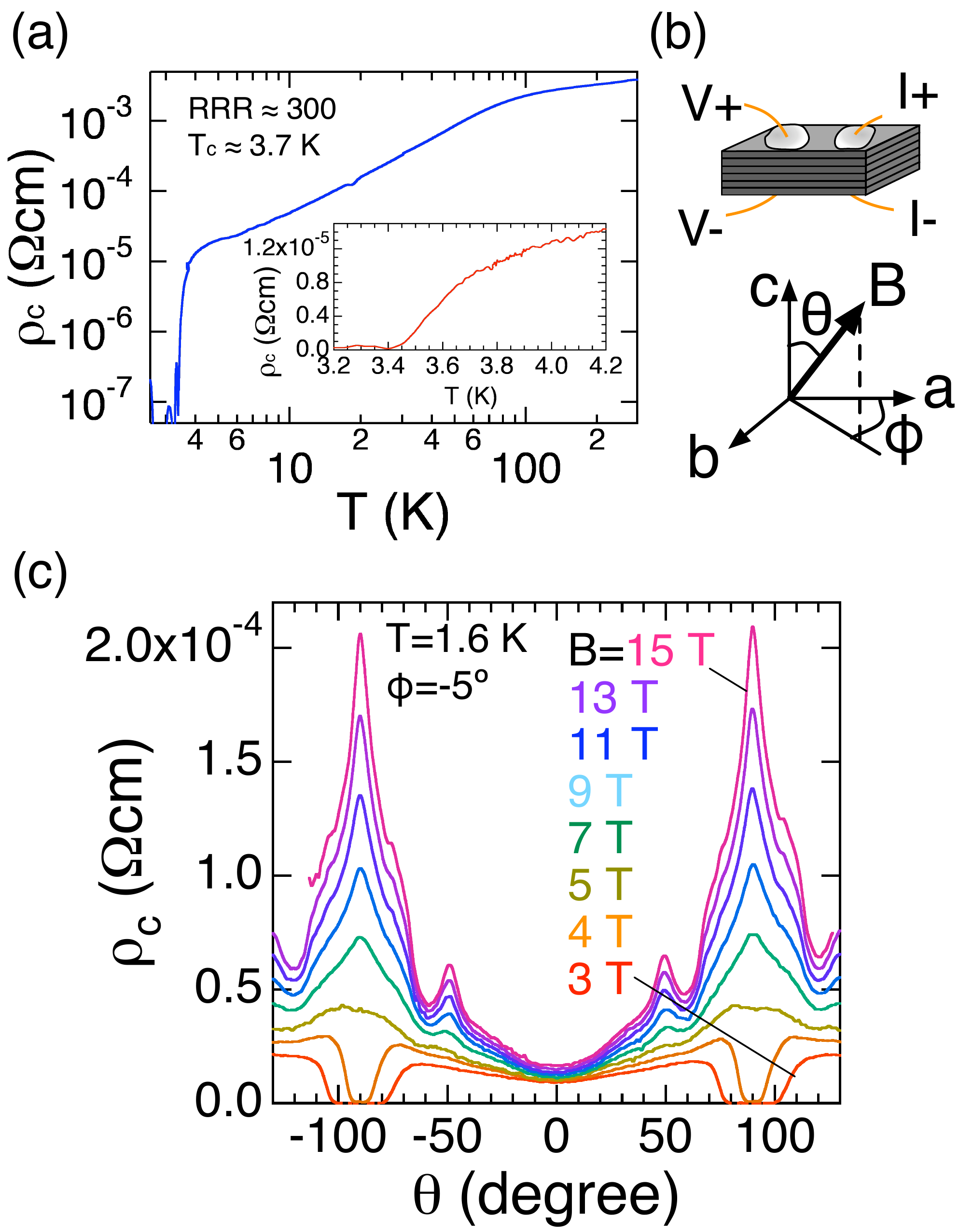}
\caption{\label{f1}(color online). (a)Temperature dependence of the $c$-axis resistivity of KFe$_2$As$_2$. The large residual resistance ratio (RRR$\approx$300) shows the high quality of the present sample. The superconducting transition temperature is about 3.7 K (inset). (b)Schematic figure of the contact configuration (top) and definition of the magnetic field angles $\theta$ and $\phi$ (bottom). (c) $\theta$ dependence of the magnetoresistance (MR) in various magnetic field strength for T=1.6 K and $\phi$=-5$^{\circ}$. Two important features are evident: small peaks at $\theta=\pm50^{\circ}$ and large peaks at $\theta=\pm90^{\circ}$. The small peaks are ascribed to AMROs arising from the geometrical effect of the Q2D FSs on the quasiparticle motion. The large peaks are ascribed to occurrence of small closed orbits on the side of Q2D FS and indicate coherent interlayer electron transport. }
\end{figure} 

[Ba$_{\rm{(1-x)}}$K$_{\rm{x}}$]Fe$_2$As$_2$ (x=0-1) is one of the most studied series among the Fe-based superconductors.The parent compound BaFe$_2$As$_2$ (x=0) has a structural transition around 140 K and no superconducting transition is observed. The superconductivity is induced by K substitution (hole doping) with x$>$0.1, and the highest $T_{\rm c}$ of 38 K is obtained for x=0.4-0.5\cite{rotter,chen}. The superconducting gap symmetry of the nearly optimally doped samples have been investigated by ARPES, microwave penetration depth, nuclear magnetic resonance, and specific heat measurements. These studies suggest that the optimally doped sample is essentially a full gap superconductor with different gap amplitudes at the multiple FSs\cite{ding, nakayama,hashimoto,fukazawa,mu,yashima}. The extremely hole-doped sample, KFe$_2$As$_2$ (x=1), also shows superconductivity with $T_{\rm c}$ of about 3.5 K\cite{rotter,chen}.  In spite of the same crystal structure, it is proposed that KFe$_2$As$_2$ is a nodal multi-gap superconductor\cite{fukazawa,dong}. These results apparently show that the FS structures and characters (electron or hole pockets), play a crucial role in the symmetry of the order parameter and probably the mechanism as well. The detailed knowledge of FS structures would lead to the overall understanding of the superconductivity in the Fe-based superconductors.

Single crystals of KFe$_2$As$_2$ were synthesized by a flux method using K as a flux. The resistivity measurements were performed by a standard four-contact method with ac current of 0.1 mA-2 mA. Four 10 $\mu$m Au wires were attached with silver paint on the cleaved sample surfaces [see Fig. 1(b)]. 

\begin{figure}
\includegraphics[width=6cm]{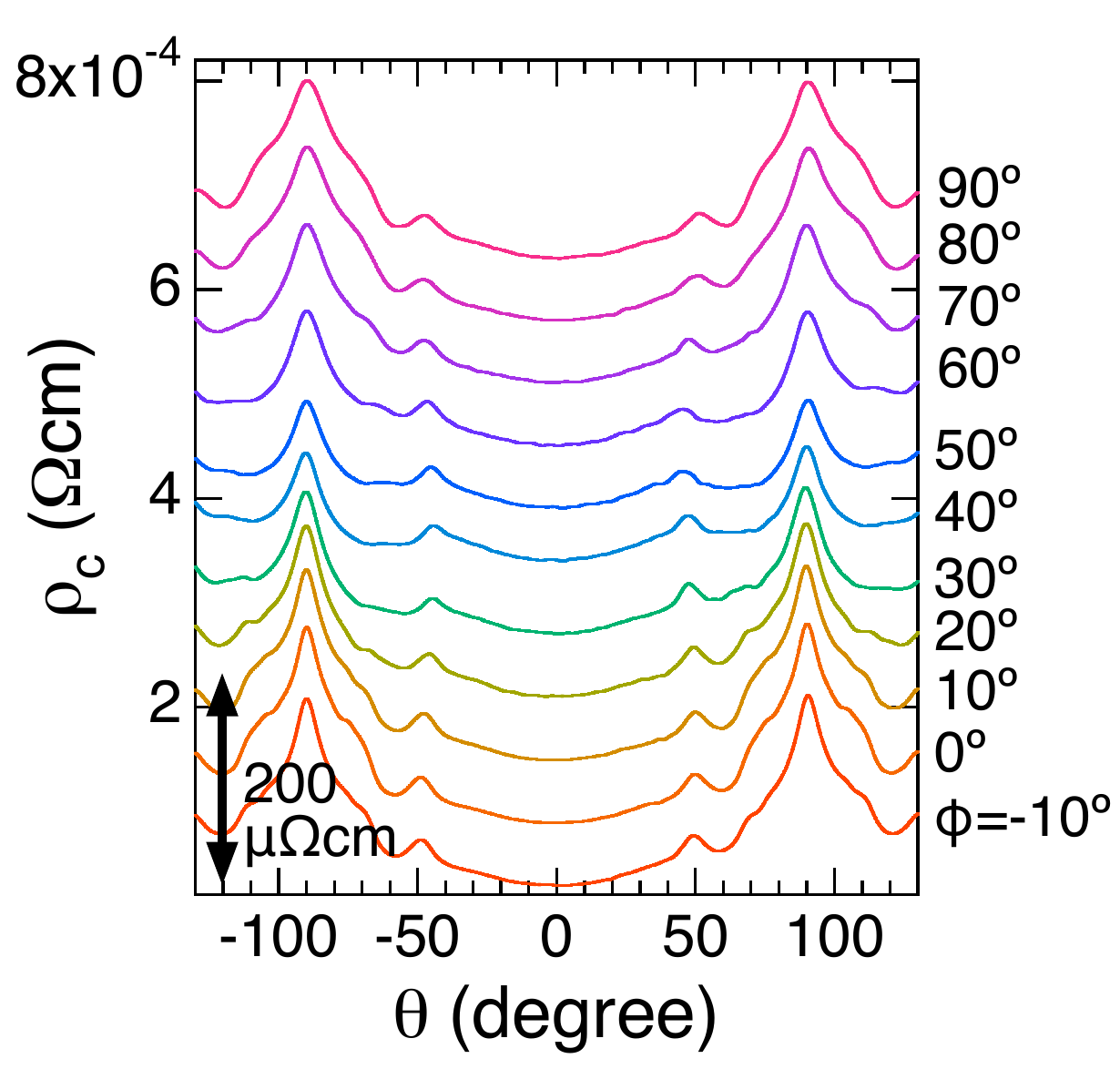}
\caption{\label{f2}(color online). $\theta$ dependence of the MR at various $\phi$. Each trace is shifted along the vertical axis for clarify. }
\end{figure}

Figure 1(a) shows the temperature dependence of the $c$-axis resistivity. The residual resistance ratio (RRR=$\rho$(300 K)/$\rho$(5 K)) amounts to about 300. This value is much higher than the previously reported value ($\sim$90)\cite{terashima}, indicating the higher quality of the sample. The superconducting transition temperature is 3.7 K [inset of Fig. 1(a)]. At 1.6 K, the superconducting state is broken by an external magnetic field of about 1 T, perpendicular to the conducting plane ($ab$-plane). Figure 1(c) shows the $\theta$ dependence of the $c$-axis MR in various fields at 1.6 K and $\phi$=-5$^{\circ}$. The definition of the magnetic field angles $\theta$ and $\phi$ are shown in the figure 1(b). We note two important features: small peaks at $\pm50^{\circ}$ and large peaks at $\pm90^{\circ}$. The small peak positions do not depend on the field strength, showing that the observed peak structures originate not from the quantum oscillations such as SdH oscillations but from AMROs. The large peaks at $\pm90^{\circ}$ will be discussed later. Superconducting transition is evident below 4T near $\theta=\pm90^{\circ}$ because of the high upper critical field parallel to the layers. Figure \ref{f2} shows the $\theta$ dependence of the $c$-axis MR with various azimuth $\phi$, and similar peak structures are clearly observed. 

AMRO in a Q2D system is interpreted in terms of Boltzmann transport theory taking into account the quasiparticle trajectories across the corrugated cylindrical FS in tilted magnetic fields.  As the field is tilted, the quasiparticle group velocity along the interlayer direction ($c$-axis for KFe$_2$As$_2$) averaged over the orbits periodically disappears, which gives interlayer resistance peaks. The resistance peak angles have a relation tan$\theta=\pi(n-\xi)/[c'k_{F}(\phi)]$ $(n=1,2,3..)$, [11] where $c'$, $k_{F}(\phi)$, and $\xi$ are the layer spacing, the Fermi wave number in the conducting plane, and the "phase factor", respectively. The phase factor $\xi$ is 1/4 if the FS warping along the least conducting direction is given by a simple cosine curve, and is isotropic in the layer. In general, $\xi$ has different values depending on the higher order corrugation of the Q2D FS. Actually, various phase factors between 0.2 and 0.9 are observed in several Q2D systems\cite{iye}. However, the period of the AMRO peaks as a function of tan$\theta$, $\Delta(\phi)$ is not affected by $\xi$, 
\begin{eqnarray}
\Delta(\phi)=\pi/[c' k_F(\phi)]. 
\label{eq1}
\end{eqnarray}
This relation directly gives the Fermi wave number $k_F(\phi)$ from $\Delta(\phi)$. The $c$-axis MR and the second derivative curve ($-d^2\rho_c/d\theta^2$) vs tan$\theta$ for $\phi=-5^{\circ}$, $-20^{\circ}$, and$-50^{\circ}$ at 1.6 K and 15 T are in Figure 3(a). In the derivative curve, the five peaks are observed. Among them, relatively large peaks denoted by B$_1$, B$_2$, and B$_3$ are periodic with tan$\theta$, so that they can be ascribed to the AMRO of a same Q2D FS cylinder. For the other small peaks (A$_1$ and A$_2$), we assume that they originate from another Q2D cylinder. Consequently, we obtain two Q2D FSs from the A and B series, whose polar plots of $k_F(\phi)$ are shown in the Fig. 3(b). Here, we take c'=c/2=6.935 ${\rm \AA}$ because the unit cell contains two layers.  We note that the cross sections of the two FSs have rounded-square shape with large $k_F$ along the $k_a$-axis. The cross sectional areas are about 12 and 17\% of the first Brillouin zone (FBZ). Because of the small numbers of the AMRO peaks and the broad ones, the accuracy of the peak period is rather limited: we estimate the error in $k_F$ to be $\sim\pm10\%$ and hence that of the FS areas to be $\sim\pm20\%$. The rounded-square shapes of the FSs are also confirmed by the $\phi$ dependence of the resistance for $\theta=\pm90^{\circ}$ [Fig. 4(a)]. A theoretical analysis\cite{morinari} indicates that, when the magnetic field is parallel to the conducting layers, dominant contributions to interlayer conductivity come from the FS parts where the Fermi velocity is parallel to the magnetic field. The resistance is therefore expected to take minima at $\phi=45^{\circ}$ in the case of the FSs shown in Fig. 3(b), which is indeed observed experimentally.

The observed total volume (12 + 17\%) is smaller than the expected one (50\%) from the carrier concentration: there exists a missing FS, whose AMRO amplitude is quite small by some reason.  The necessary condition for the AMRO observation is roughly $\omega_c\tau>$1, where $\omega_c$ and $\tau$ are the cyclotron frequency and the scattering time, respectively.  Since $\omega_c$ is inversely proportional to the effective mass $m_c$, the missing FS likely has large $m_c$ and/or shorter $\tau$. 

The initial ARPES study reported two Q2D FSs, $\alpha$ and $\beta$, around the $\Gamma$, whose cross sections were 7 and 22\% of the BZ, respectively \cite{sato}.  A more recent study \cite{yoshida} has revealed that the $\alpha$ FS is actually quasi-degenerate. The obtained cross sections are 10.1, 11.8, and 28.5\% of the BZ for the inner and outer FSs of the quasi-degenerate $\alpha$ and the $\beta$ FSs, respectively. Our dHvA study has also found two Q2D FSs, $\alpha$ and $\zeta$, with the cross sections of 8.4 and 13\% of the BZ \cite{terashima2}.  They correspond to the inner and outer FSs of the quasi-degenerate $\alpha$ FS.  Effective masses of electrons on these FSs are in a range from 6 to 18m$_e$. The size of the unobserved $\beta$ FS is calculated to be 24\% of the BZ, and with the help of single crystal specific heat data \cite{fukazawa3}, the associated effective mass can be estimated to be 18m$_e$ on average, heavier than the average masses of $\alpha$ and $\zeta$ (6.3$m_e$ and 13$m_e$, respectively). Considering the 20\% error of the present area estimations, we identify the presently observed two FSs with the $\alpha$ and $\zeta$ FSs of the dHvA studies.

\begin{figure}
\includegraphics[width=7.3cm]{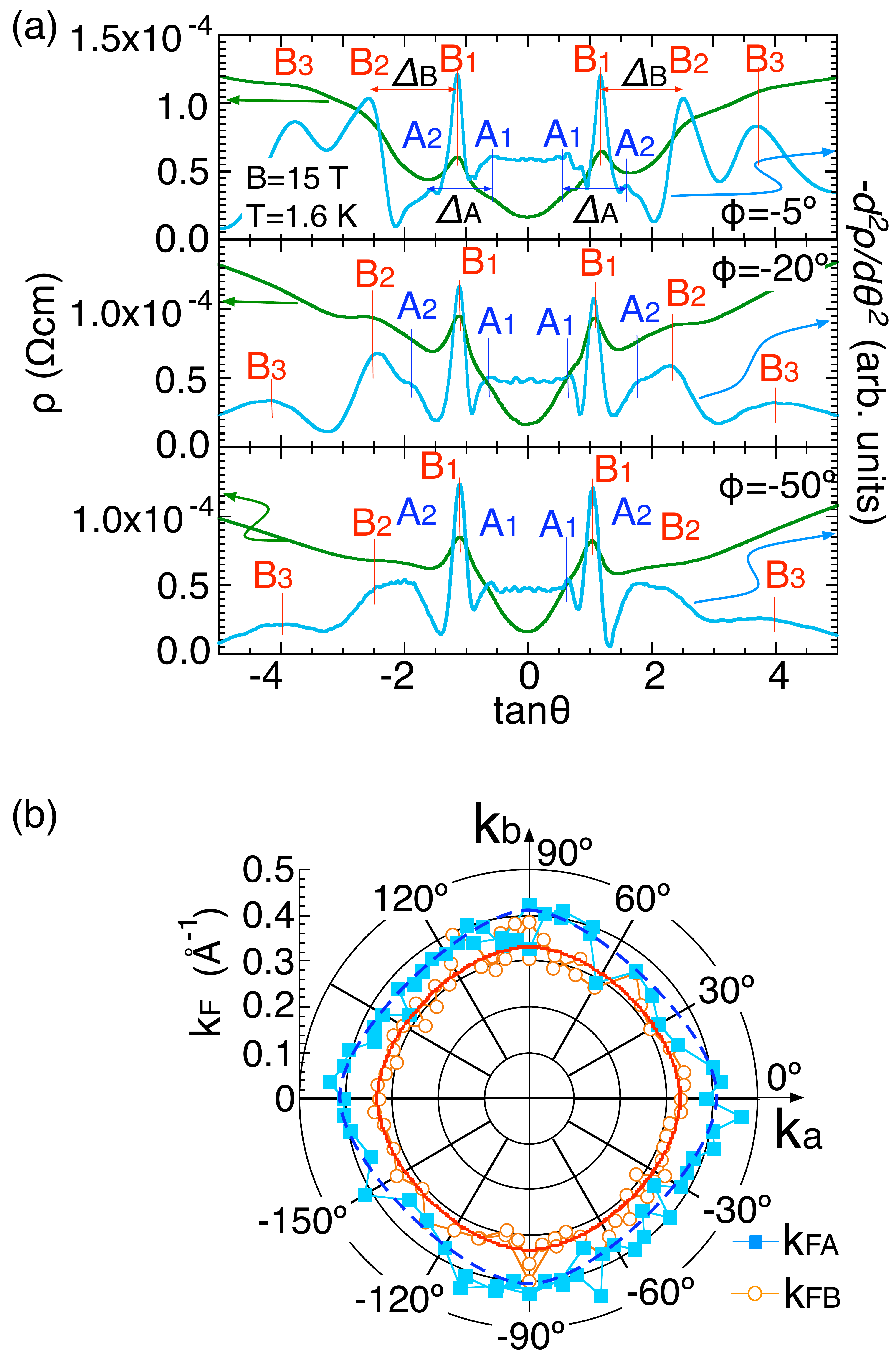}
\caption{\label{f3}(color online). (a) Magnetoresistance (left axis) and the second derivative curve (right axis) as a function of tan$\theta$ at 1.6 K for B=15 T and $\phi=-5^{\circ}$, $-20^{\circ}$, $-50^{\circ}$. Relatively large and periodic peaks (B$_1$, B$_2$ and B$_3$) can be ascribed to the AMRO of a same Q2D FS cylinder, and small peaks (A$_1$ and A$_2$) likely to another.  
(b) Polar plot of the Fermi wave number ($k_F(\phi)$) obtained from $\Delta_A$ and $\Delta_B$ (see (a) in this figure). We used Eq. \ref{eq1} and $c'=c/2=6.935$ ${\rm \AA}$ to calculate $k_F(\phi)$. To obtain $k_{FB}(\phi)$, we only use the peak positions of B$_1$ and B$_2$ because they are much clearer than B$_3$. Solid and dashed lines show the calculated results with the equation $k_a^n+k_b^n=r^n$. The best fits were obtained when $r$=0.41 ${\rm \AA}^{-1}$, $n$=1.5, and $r$=0.33 $\rm \AA^{-1}$, $n$=1.7 for $k_{FA}(\phi)$ and $k_{FB}(\phi)$, respectively. The cross sections correspond to about 12 and 17$\%$ of the FBZ, respectively.}  
\end{figure}

In the above AMRO analyses, the presence of Q2D FSs is tacitly assumed, where the quasiparticles coherently move in any directions. However, a theoretical investigation reveals that the AMRO is observed even when the interlayer transfer is weakly incoherent: the electrons are scattered more frequently than they tunnel between the layers\cite{moses}. One of the decisive tests for the interlayer coherent transport is the observation of the interlayer MR peak with field-independent peak width $\delta\theta$ under sufficiently high fields parallel to the layers\cite{hanasaki}. Figure 4(b) shows the first derivative curves of the $c$-axis MR around $\theta=-90^{\circ}$. The peak widths of the MR curves, which are defined as peak-to-peak widths in the first derivatives (indicated by the dashed lines), are almost field independent. The similar features are observed at any azimuths $\phi$. The results clearly lead to the conclusion that the interlayer transport is coherent in KFe$_2$As$_2$. The peak width $\delta\theta$ is proportional to interlayer transfer integral $t_c$ as follows\cite{hanasaki}, 
\begin{eqnarray}
\delta\theta\approx2k_F t_c c'/E_F,  
\label{eq2}
\end{eqnarray}
where $E_F$ is Fermi energy. From Eq.\ref{eq2}, we obtain $t_c/E_F\approx0.03$ for $\delta\theta=6^{\circ}$. Such small $t_c/E_F$ is consistent with the two dimensionality of the $\alpha$ and  $\zeta$ orbits. Indeed, from the dHvA results\cite{terashima2}, the $t_c/E_F$ estimated for the $\alpha$ and $\zeta$ orbits are 0.01 and 0.1, respectively.

\begin{figure}
\includegraphics[width=8.5cm]{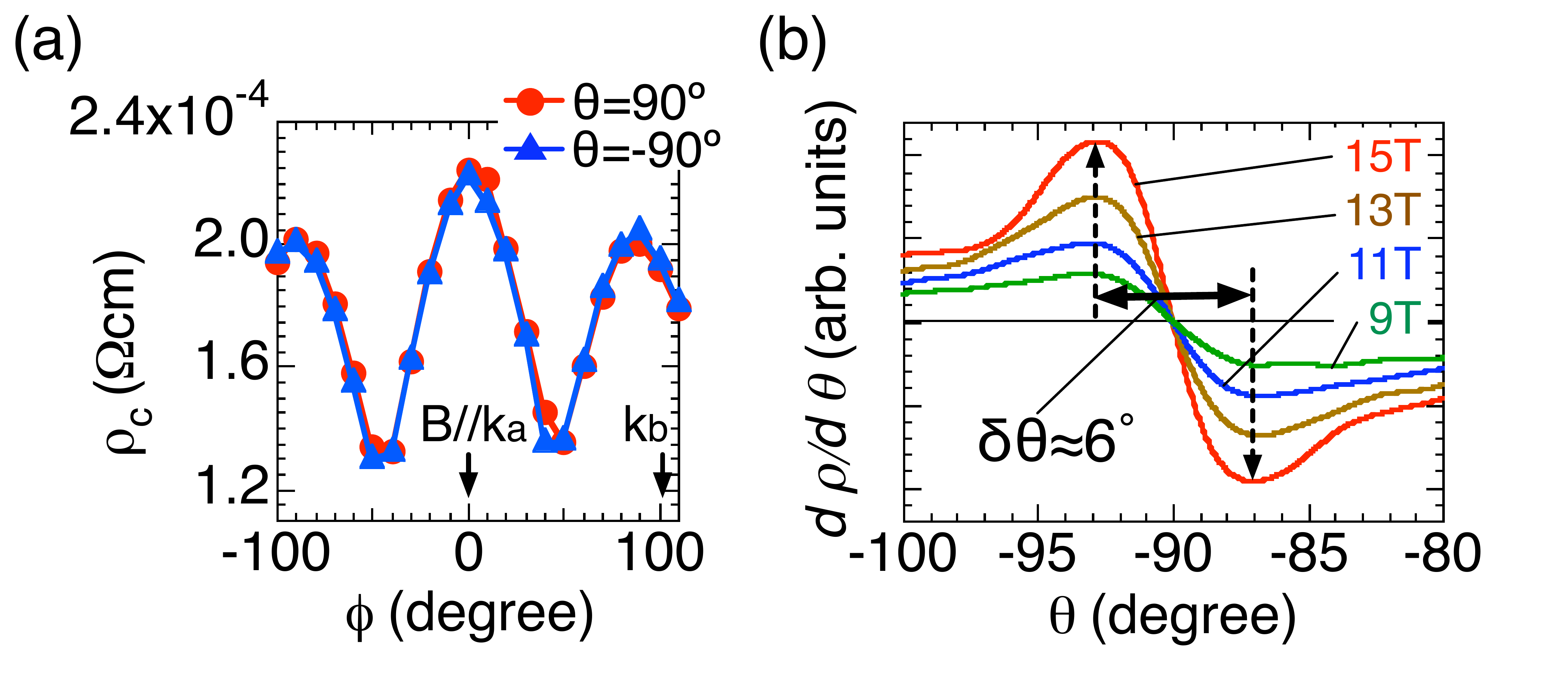}
\caption{\label{f4}(color online). (a) $\phi$ dependence of the interlayer resistivity in inplane field.  The maxima at $\phi=0^{\circ}, \pm90^{\circ}$ and minima at $\phi=\pm45^{\circ}$ are consistent with the rounded square shape of the FSs shown in Fig. 3(b)\cite{morinari}. (b) The first derivative curves of c-axis MR around $\theta=-90^{\circ}$ at different fields for $\phi=-5^{\circ}$. The peak width of the MR, defined as the peak-to-peak width of the derivative, (shown by dashed lines) is almost independent of field. } 
\end{figure}

At present, no first principle band calculation can reasonably reproduce the precise topology of the FSs of KFe$_2$As$_2$. However, the rounded-square shape of the FSs can be partially understood by a tight binding approximation of the Fe $3d$ orbitals. Because of the ligand field of the Fe ions, the two kinds of $3d$ orbitals $d_{XZ}$ and $d_{YZ}$ are doubly degenerated. Here, we have defined the orthogonal nearest neighbor Fe-Fe directions as $X$ and $Y$ direction, respectively. Since the $d_{XZ}$ ($d_{YZ}$) orbitals overlap along the $X$ ($Y$) direction much more than along the $Y$ ($X$) direction, they would form a quasi-one-dimensional FS perpendicular to the $X$ ($Y$) direction if hybridization between the $d_{XZ}$ and $d_{YZ}$ orbitals neglected. When the hybridization is turned on, a rounded-squared shape FS would result around the $\Gamma$ point. The $3d$ orbitals, elongated in the interlayer direction, would give relatively large transfer integral in the interlayer direction. This is consistent with the coherent interlayer transport.

In order to understand the mechanism of the high T$_c$ of the Fe-based superconductors, theories on strongly correlated electrons (beyond BCS theory) are apparently required. A possible mechanism is antiferromagnetic (AF) fluctuation between the different FS pockets as has been discussed in many literatures, e.g., \cite{mazin1,kuroki,mazin2}.  The AF fluctuation is more enhanced in low dimensional electronic structure, because it is closely related to the nesting instability of the FS. Our data show the presence of multiple Q2D FSs with rounded square cross sections. Consequently, the FSs have rather flat parts, which make the nesting instability high. These results are reasonably consistent with many theories, which argue that the AF fluctuation plays a crucial role in the superconductivity mechanism of the Fe-based superconductors. In KFe$_2$As$_2$, the main FSs are located at the $\Gamma$ point \cite{sato, yoshida}. When the AF fluctuation between the flat part of the same main FS and/or the different ones is the dominant mechanism of the superconductivity, we expect the presence of the line node of the order parameter, which is consistent with the specific heat and the thermal conductivity measurements \cite{fukazawa,dong}.

In conclusion, we have determined the FSs in an extremely hole doped Fe-based superconductor KFe$_2$As$_2$ by a bulk probe, AMROs measurements. From the two series of the AMROs, two Q2D FS cylinders are derived, whose cross sectional areas correspond to 12 and 17\% of the FBZ.
Both the cross sections have a rounded-square shape with the long axes along the $k_a$ and $k_b$ directions. This can be partially understood in terms of the hybridization of the Fe $d_{XZ}$ and $d_{YZ}$ orbitals. The character of the orbitals suggests some finite amount of the interlayer hybridization. It is consistent with the observation of the sharp peaks in the interlayer resistance under the in-plane field: the interlayer transport is coherent. So far agreement between experimental and theoretical FSs in the iron pnictides is rather limited \cite{coldea, terashima2}, and the roles of the electronic correlations are under debate\cite{ortenzi,Aichhorn,ikeda}. The present determination of the FS shapes with suggestion of the mainly contributing orbital characters will help our understanding of electronic structures and the superconducting mechanisms in the iron pnictides.

This work was supported by Grant-in-Aid for Scientific Research on Innovative Areas, (Nos. 20110004 and 20102005) from JSPS of Japan.


\begin{thebibliography}{99}

\bibitem{kamihara}
Y. Kamihara, {\it et al}., J. Am. Chem. Soc {\bf 130}, 3296 (2008). 
\bibitem{kito}
H. Kito, {\it et al}., J. Phys. Soc. Jpn {\bf 77}, 063707 (2008).
\bibitem{ren}
Z.-A. Ren, {\it et al}., Chin. Phys. Lett. {\bf 25}, 2215 (2008).
\bibitem{lebegue}
S. Leb$\grave{\rm e}$gue, Phys. Rev. B {\bf 75}, 035110 (2007).
\bibitem{shingh}
D. J. Singh, Phys. Rev. B {\bf 79}, 174520 (2009).
 \bibitem{coldea}
 I. Coldea, {\it et al}., Phys. Rev. Lett. {\bf 101}, 216402 (2008).
\bibitem{lu}
D. H. Lu, {\it et al}., Nature {\bf 455}, 81 (2008). 
\bibitem{sato}
T. Sato, {\it et al}., Phys. Rev. Lett. {\bf 103}, 047002 (2009).
\bibitem{kar}
M. V. Kartsovnik, {\it et al}., Pis'ma Zh. Eksp. Teor. Fiz {\bf 48}, 498 (1989) [JETP Lett., {\bf 48}, 541 (1988)]. 
\bibitem{kajita}
K. Kajita, {\it et al}., Solid State Commun., {\bf 70}, 1181 (1989). 
\bibitem{yamaji}
K. Yamaji, J. Phys. Soc. Jpn., {\bf 58}, 1520 (1989).
\bibitem{rotter}
M. Rotter, {\it et al}., Phys. Rev. Lett. {\bf 101}, 107006 (2008). 
\bibitem{chen}
H. Chen, {\it et al}., Europhys. Lett. {\bf 85}, 17006 (2009). 
\bibitem{ding}
H. Ding, {\it et al}., Europhys. Lett. {\bf 83}, 47001 (2008). 
\bibitem{nakayama}
 K. Nakayama, {\it et al}., Europhys. Lett. {\bf 85}, 67002 (2009). 
\bibitem{hashimoto}
K. Hashimoto, {\it et al}., Phys. Rev. Lett. {\bf 102}, 207001 (2009). 
\bibitem{fukazawa}
H. Fukazawa, {\it et al}., J. Phys. Soc. Jpn. {\bf 78}, 033704 (2009). 
\bibitem{mu}
G. Mu, {\it et al}., Phys. Rev. B {\bf 79}, 174501 (2009). 
\bibitem{yashima}
M. Yashima, {\it et al}., J. Phys. Soc. Jpn. {\bf 78}, 103702 (2009). 
\bibitem{fukazawa2}
H. Fukazawa, {\it et al}., J. Phys. Soc. Jpn. {\bf 78}, 083712 (2009).  
\bibitem{dong}
J.K. Dong, {\it et al}., Phys. Rev. Lett. {\bf 104}, 087005 (2010). 
\bibitem{terashima}
T. Terashima, {\it et al}., J. Phys. Soc. Jpn. {\bf 78}, 063702 (2009). 
\bibitem{iye}
Y. Iye, {\it et al}., J. Phys. Soc. Jpn. {\bf 63}, 1643 (1994); 
Y. Iye, Materials Science and Engineering {\bf B31}, 141 (1995).  
\bibitem{morinari}
T. Morinari, {\it et al}., J. Phys. Soc. Jpn. {\bf 78}, 114702 (2009). 
\bibitem{moses}
P. Moses, and R. H. McKenzie, Phys. Rev. B {\bf 60}, 7998 (1999). 
\bibitem{hanasaki}
N. Hanasaki, {\it et al}., Phys. Rev. B {\bf 57}, 1336 (1998).  
\bibitem{yoshida}
T. Yoshida, {\it et al}., arXiv:1007.2698
\bibitem{fukazawa3}
H. Fukazawa, private communication. 
\bibitem{terashima2}
T. Terashima, {\it et al}., J. Phys. Soc. Jpn. {\bf 79}, 053702 (2010).


\bibitem{mazin1}
I. I. Mazin, {\it et al}., Phys. Rev. Lett. {\bf 101}, 057003 (2008).
\bibitem{kuroki}
K. Kuroki, {\it et al}., Phys. Rev. Lett. {\bf 101}, 087004 (2008). 
\bibitem{mazin2}
I. I. Mazin and J. Schmalian, Physca C {\bf 469}, 614 (2009).
\bibitem{ortenzi}
L. Ortenzi, {\it et al}., Phys. Rev. Lett., {\bf 103}, 046404 (2009). 
\bibitem{Aichhorn}
M. Aichhorn, {\it et al}.,  Phys. Rev. B {\bf 80}, 085101 (2009). 
\bibitem{ikeda}
H. Ikeda, {\it et al}., Phys. Rev. B {\bf 82}, 024508 (2010). 

\end{thebibliography}

\end{document}